\documentclass[iop]{emulateapj}
\usepackage{apjfonts}
\usepackage{amsmath}
\usepackage{graphicx}
\usepackage{natbib}
\usepackage{multirow}
\usepackage{mathrsfs} %curl fonts

\def \na {New Astron.}

\shorttitle{Understanding Compact Object formation. IV.}
\shortauthors{T.-W. Wong et al.}

\begin{document}

\title{Understanding Compact Object Formation and Natal Kicks. IV. The case of IC 10 X-1}
\author{Tsing-Wai Wong$^{1,2}$,
Francesca Valsecchi$^1$,
Asna Ansari$^{1,3}$,
Tassos Fragos$^2$,
Evert Glebbeek$^4$,
Vassiliki Kalogera$^1$,
Jeffrey McClintock$^2$
}

\affil{
$^1$Center for Interdisciplinary Exploration and Research in Astrophysics (CIERA) \& Department of Physics and Astronomy, Northwestern University,
\\2145 Sheridan Road, Evanston, IL 60208, USA; tsingwong2012@u.northwestern.edu, francesca@u.northwestern.edu,vicky@northwestern.edu
\\$^2$Harvard-Smithsonian Center for Astrophysics, 60 Garden St., Cambridge, MA 02138, USA; tfragos@cfa.harvard.edu, jem@cfa.harvard.edu
\\$^3$Lamont-Doherty Earth Observatory, Columbia University, Palisades, NY 10964, USA; ansari@ldeo.columbia.edu
\\$^4$Department of Astrophysics/IMAPP, Radboud University Nijmegen, PO Box 9010, 6500 GL, Nijmegen, The Netherlands; e.glebbeek@astro.ru.nl
}

%%%%%%%%%%%%%%%%%
%%% Beginning of Abstract %%%
%%%%%%%%%%%%%%%%%
\begin{abstract}
The extragalactic X-ray binary IC 10 X-1 has attracted attention as it is possibly the host of the most massive stellar-mass black-hole (BH) known to date.
Here we consider \emph{all} available observational constraints and construct its evolutionary history up to the instant just before the formation of the BH.
Our analysis accounts for the simplest possible history that includes three evolutionary phases: binary orbital dynamics at core collapse, common envelope (CE) evolution,
and evolution of the BH--helium star binary progenitor of the observed system.
We derive the complete set of constraints on the progenitor system at various evolutionary stages.
Specifically: right before the core collapse event, we find the mass of the BH immediate progenitor to be $\gtrsim 31~M_\sun$ (at 95\% of confidence, same hereafter).
The magnitude of the natal kick imparted to the BH is constrained to be $\lesssim 130$~km/s.
Furthermore, we find that the ``enthalpy'' formalism recently suggested by Ivanova $\&$ Chaichenets is able to explain the existence of IC\;10\;X-1 without the need of invoking
unreasonably high CE efficiencies.
With this physically motivated formalism, we find that the CE efficiency required to explain the system is in the range of $\simeq 0.6$--1. 
\end{abstract}


%%%%%%%%%%%%%%%%%%%
%%% Beginning of Introduction %%%
%%%%%%%%%%%%%%%%%%%
\section{Introduction}

Over the past few decades, it has become clear that neutron stars (NS) receive recoil kicks at birth (also known as natal kicks) during the core collapse event.
This conclusion is based on proper motion studies of pulsars \citep[see e.g.][]{gunn-ostriker70, lyne...82, lyne-lorimer94, brisken...03, hobbs...05, chatterjee...09}
and evolutionary studies of NS-hosting binaries \citep[see e.g.][]{brandt-podsiadlowski95, pfahl...02, thorsett...05, willems...06, martin...09, wong...10}.
However, whether black holes (BH) receive similar natal kicks during the core collapse event is still uncertain \citep[see e.g.][]{brandt...95, nelemans...99, gualandris...05, dhawan...07, repetto...12}.
If BH kicks are ubiquitous, then BH formation must be closely associated to that of NS before the formation of the event horizon. Otherwise, the formation of BHs through more than
one physical process will be favored.

To shed light on questions related to BH formation, we perform detailed evolutionary modeling of the observed BH X-ray binaries (XRB), which enables us to derive robust constraints
on the mass of the BH immediate progenitor and the magnitude of the natal kick imparted to BH during the core collapse event. In the previous three papers of this series, we studied the XRB
GRO J1655--40 \citep{willems...05}, XTE J1118+480 \citep{fragos...09} and Cygnus X-1 \citep{wong...12}.
For GRO J1655--40, we constrained the mass of the BH immediate progenitor to be $\simeq 5.5$--11.0~$M_\sun$ and the magnitude of its natal kick to be $\lesssim 210$~km/s.
In the study of J1118+480, we found that a natal kick of magnitude $\simeq 80$--310~km/s is required to explain the formation of this system,
and also derived a lower limit of $\simeq 6.5$~$M_\sun$ on the mass of the BH immediate progenitor.
Finally, for Cygnus X-1 we constrained the mass of the BH immediate progenitor and the magnitude of its natal kick to be $\gtrsim 15 M_\sun$ and $\lesssim 77$~km/s, respectively.
Similarly, \cite{ValsecchiNature2010} studied the formation of M33 X-7 using binary modeling as well and derived the mass of the BH immediate progenitor and the magnitude of the natal
kick to be in the range of $\simeq 15.0$--16.1~$M_\sun$ and $\simeq 10$--850~km/s, respectively. In this paper, we study the formation of the BH in the extragalactic XRB IC 10 X-1.

IC 10 X-1 is one of the four observed BH XRBs that are known to host a Wolf-Rayet (WR) star as the mass donor \citep{clark-crowther04}.
The other three systems are Cygnus X-3 \citep[see e.g.][]{zdziarski...13}, NGC 300 X-1 \citep[see e.g.][]{crowther...10}, and M 101 ULX-1 \citep[see e.g.][]{liu...13}.
The X-ray emission is powered by the accretion of stellar wind material onto the BH.
At the present time, the BH in IC 10 X-1 is the most massive known stellar-mass BH \citep[$\simeq 23$--34~$M_\sun$,][]{prestwich...07, silverman-filippenko08}.
Since the supergiant progenitor of the observed WR star cannot fit into the tight orbit at the present time \citep[orbital period~=~34.93~hr,][]{silverman-filippenko08},
it is natural to consider the system's evolution via a common envelope (CE) evolution phase, which involves the BH and the massive progenitor of the observed WR star.
However, such a binary is likely to merge at the end of the CE phase, as the envelopes of massive stars are tightly bound \citep{podsiadlowski...03}.

\cite{demink...09} suggested an alternative formation scenario for IC 10 X-1 that does not invoke any CE phase, which they called ``Case $M$'' evolution.
They considered two massive stars in a tight orbit, such that tidal interactions always kept them spinning rapidly.
This led to efficient mixing of elements throughout their stellar interiors via rotational effects, and hence both stars went through chemical homogenous evolutions.
They stayed compact throughout their main sequence evolution and turned into abnormally massive helium (He) stars without going through any CE evolution. Although Case $M$ helps to
explain the high masses of the BH and the WR star in IC 10 X-1, the short orbital period gives rise to a difficulty: the intense mass loss suffered by the WR star and its progenitor will widen
the orbit. It is very hard to explain the current tight orbit without a CE phase. As a result, the evolution history of IC 10 X-1 has remained uncertain.

Instead of studying its past evolution, \cite{bulik...11} performed binary modeling to predict the fate of IC 10 X-1.
They estimated that the observed WR star will go through core collapse in $\lesssim 0.3$~Myr, leading to the formation of a close double BH binary with a short coalescence time
($\sim 3$~Gyr). 

The plan of the paper is as follows. In \S\ref{sec:obs_constraints}, we review the current available observational constraints of IC 10 X-1. A general outline of our analysis methodology is
presented in \S\ref{sec:methodology}, while detailed discussions of each individual step are in \S\ref{sec:postCE_binary_modeling}--\ref{sec:dynamics@CC}. Our derived constraints
related to the formation of the BH and the past evolution of IC 10 X-1 are discussed in \S\ref{sec:results}, and we offer our conclusions in the final section.

%%%%%%%%%%%%%%%%%%%%%%%%
%%% Beginning of Observations Section  %%%
%%%%%%%%%%%%%%%%%%%%%%%%
\section{Observational Constraints for IC 10 X-1}
\label{sec:obs_constraints}

IC 10 X-1 is a persistent X-ray source in the local starburst galaxy IC 10. It was discovered by \cite{brandt...97} in their X-ray observations of IC 10 made with ROSAT.
\cite{bauer-brandt04} derived the 0.5--8.0 keV unabsorbed X-ray luminosity of IC 10 X-1 to be $1.50 \times 10^{38}$ ergs/s from their Chandra observations. 
The optical counterpart of IC 10 X-1 was identified as the WR star [MAC92] 17A by \cite{clark-crowther04}.
Based on the \cite{schaerer-maeder92} mass-luminosity relationship, they estimated the mass of this WR star to be 35 $M_\sun$.
Using the data from Chandra and Swift observations, \cite{prestwich...07} determined the orbital period of IC 10 X-1 to be $34.40 \pm 0.83$ hr and obtained a mass function
of 7.8 $M_\sun$. Using their mass function and considering an inclination angle of 90 degrees, they estimated the mass of the BH to be 23--34 $M_\sun$.
Their results indicated that IC 10 X-1 hosted the most massive stellar mass BH known at the present time. The observed orbital period also implied that the BH is currently accreting mass
from the intense stellar wind of its WR companion. \cite{silverman-filippenko08} precisely measured the radial velocity amplitude of the WR star and refined the orbital period and mass
function to $34.93 \pm 0.04$~hr and $7.64 \pm 1.26$~$M_\sun$, respectively. For convenience, our adopted observational constraints are summarized in Table~\ref{tab:observed_properties}.

\begin{deluxetable*}{lccc}
\tablewidth{18.0 cm}
\tablecolumns{4}
\tablecaption{Properties of IC 10 X-1}
\tablehead{
\colhead{Parameter} & \colhead{Notation} & \colhead{Value} & \colhead{References}
}
\startdata
Distance (kpc) & $d$ & $590 \pm 35$ & (1), (3)\\
Orbital period (hr) & $P_{orb}$ & $34.93 \pm{0.04} $ & (6)\\
Inclination angle (deg) & $i$ & 65--90 & (6)\\
Mass Function ($M_\sun$) & $f(M_{BH}$) & $7.64 \pm 1.26$ & (6)\\
Black hole mass ($M_\sun$) & $M_{BH}$ & $\simeq$23--34 & (5)\\
Companion mass ($M_\sun$) & $M_2$ & 17--35 & (3), (5)\\
Companion Luminosity ($10^6~L_\sun$) & $L_2$ & $1.122 \times (d / 590~\rm kpc)^2$ & (3)\\
Companion Effective temperature (K) & $T_{eff2}$ & 85 000 & (3)\\
0.5--8.0 keV unabsorbed X-ray luminosity ($10^{38}$ erg s$^{-1}$) & $L_X$ & $1.50 \times (d / 700~\rm kpc)^2$ & (2)\\
Metallicity ($Z_\sun$) & Z & 0.2 & (4)
\enddata
\label{tab:observed_properties}
\tablerefs{(1) Borissova et al. 2000; (2) Bauer \& Brandt 2004; (3) Clark \& Crowther 2004; (4) Leroy et al. 2006; (5) Prestwich et al. 2007; (6) Silverman \& Filippenko 2008.}
\end{deluxetable*}

%%%%%%%%%%%%%%%%%%%%%%%%
%%%  Beginning of Methodology Section  %%%
%%%%%%%%%%%%%%%%%%%%%%%%
\section{Outline of Analysis Methodology}
\label{sec:methodology}

Among the evolutionary scenarios that could potentially explain the formation of IC 10 X-1, we adopt the simplest possible evolutionary history that provides a consistent explanation
of \emph{all} observational constraints.
First of all, we assume that the BH progenitor and its companion were born at the same time.
Towards the end of the BH progenitor's life, it lost its hydrogen (H) rich envelope because of mass loss via a stellar wind or binary interactions.
Hence, the BH immediate progenitor is a helium (He) rich star.
Soon after the birth of the BH, the companion star evolved off the main sequence and became a supergiant.
Eventually, the star overfilled its Roche lobe and underwent a phase of dynamically unstable mass transfer, which inevitably led to CE evolution.
During this CE phase, the BH was engulfed into the H-rich envelope of its companion.
The CE phase ended when the envelope was ejected, leaving behind a binary consisting of a BH and a He star in
a tight, circular orbit. The intense mass loss via stellar wind from the He star continuously drove the binary components further away from each other.
A very small fraction of this stellar wind material was accreted onto the BH, resulting in X-ray emission from IC 10 X-1 that we see today.

This is the simplest possible evolutionary channel one can envision given the current properties of the system.
Therefore in this paper, we restrict ourselves to the formation of IC 10 X-1 through the above evolutionary channel.
In order to derived constraints on the formation of the BH in IC 10 X-1, we track the evolutionary history of the system back to the instant just before the core collapse event.
Our analysis incorporates a number of calculations that can be grouped in three main steps.
Hereafter, we add the prefix ``pre-'' and ``post-'' to the name of any event occurred in the evolutionary history of IC 10 X-1, in order to indicate the instant just before and right after
that event, respectively.
 
In the first analysis step, our goal is to derive the post-CE binary and stellar properties of IC 10 X-1.
We start by constructing He star models that satisfy the current mass and luminosity constraints given in Table~\ref{tab:observed_properties}.
Using the properties of our He star models and different BH masses and post-CE binary semi-major axes, we evolve the post-CE binary's orbit forward in time until the current epoch.
This calculation accounts for tides, wind mass loss, wind accretion onto the BH and orbital angular momentum loss via gravitational radiation. We examine
the evolutionary sequence of every binary model to find whether at the present time the BH mass and the orbital period simultaneously match the observational data,
in which case we classify that evolutionary sequence as ``successful''.
Furthermore, we consider $1\sigma$ of observational uncertainties when matching the properties of our models with the observational data.
The post-CE binary properties can then be obtained from our successful evolutionary sequences.

In the second analysis step, we study the CE event to determine whether our derived properties of the post-CE binary can be achieved. To do so we employ the standard
$\alpha$-prescription \citep{webbink84}, examining also several alternative formulation of the energy budget prescription, to compute the CE efficiency, $\alpha_{CE}$.
If this parameter is $\le 1$, we conclude that our derived properties of the post-CE binary can be explained by the current understanding of CE evolution.

In the final analysis step, we utilize the results from the first two steps and perform a Monte Carlo simulation of the orbital dynamics involved in the core collapse event.
Our goal is to derive limits and construct probability distribution functions (PDF) for the BH immediate progenitor mass and the potential natal kick magnitude imparted to the BH.
We start with randomized properties of the pre-core collapse (pre-SN) binary, which include BH immediate progenitor mass, orbital semi-major axis, eccentricity and phase.
The core collapse event is approximated as occurring instantaneously, with the mass ejection and possible asymmetries in the explosion
imparting to the BH a potential kick that has random magnitude and direction.
Using the equations of orbital energy and angular momentum, we determine whether the binary can survive the core collapse event.
If it does, we map the pre-SN binary properties to the post-core collapse (post-SN) phase space.
Then, we apply the constraint related to CE evolution and retain the data points with $\alpha_{CE} \le 1$. At the end, our constraints related to
the BH formation in IC 10 X-1 are derived from the retained data points. 

To build the stellar models necessary for our analysis, we use a version of the STARS code originally developed by Peter Eggleton 
\citep{eggleton71, eggleton73, pols...95, eggleton-eggleton02, glebbeek...08}.
The adopted opacity tables combine the OPAL opacities from \cite{iglesias-rogers96}, the low temperature molecular opacities from \cite{ferguson...05},
the electron-conduction opacities from \cite{cassisi...07} and the Compton scattering opacities from \cite{buchler-yueh76}.
The assumed heavy element composition in each stellar model is scaled according to the solar abundances described by \cite{anders-grevesse89}.
Convective boundaries are determined using the Schwarzschild criterion.
Mixing of elements due to convection and semi-convection are taken into account \citep{eggleton73, langer91}, but the effects of rotational mixing and meridional circulation are excluded.
We notice that these effects can be very important. However, their efficiencies are poorly constrained and need to be tuned using the observational data.
We choose not to add extra free parameters because our models are intended to be the simplest ones able to satisfy the observational constraints.
For mass loss during the main sequence evolution, we adopt the prescriptions of \cite{vink...01}.
If the surface H abundance drops below 0.4, we switch to the WR prescription developed by \cite{nugis-lamers02} with the metallicity scaling law determined by \cite{vink-dekoter05}.
When the effective temperature falls below the working regime (i.e. $<10$~kK) of the mass loss prescriptions developed by \cite{vink...01}, we use the mass loss rates determined by
\cite{dejager...88} instead.
This stellar-wind mass loss recipe is extrapolated into the post-main sequence evolution, including the regime of luminous blue variables (LBV).
Indeed, all of the above mass loss prescriptions give a high mass loss rate in the LBV regime,
with a range of $10^{-5}$--$10^{-2}\,M_\odot/yr$ for the massive stars involved in this analysis.
This range of rates is consistent with the measured mass loss rates of LBVs \citep[see e.g.][]{stahl...90, stahl...01, leitherer...94, hillier...01, vink-dekoter02, umana...05, mehner...13}.
In addition, there is no prescription included to simulate the mass loss occurred in giant eruptions of LBV stars.
Stellar evolutionary codes generally approximate this episodic mass loss with the averaged continuous mass loss.
We note that the mass loss rates of massive stars, especially LBVs, are poorly constrained at the present time.
The effect of this uncertainty on our study of the required CE evolution will be discussed in \S\,\ref{sec:CE_massloss}.

%%%%%%%%%%%%%%%%%%%%%%%%%%
%%%  Beginning of Post-CE Binary Modeling  %%%
%%%%%%%%%%%%%%%%%%%%%%%%%%
\section{Post-Common Envelope Binary Modeling}
\label{sec:postCE_binary_modeling}

\begin{figure}
\epsscale{1.0}
\plotone{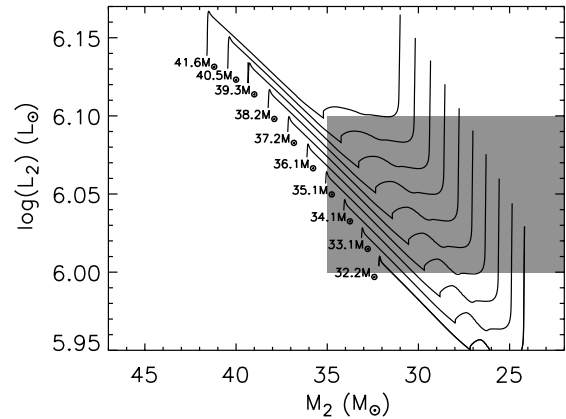}
\caption{
Luminosity as a function of mass for He star models with metallicity Z = 0.004 and initial masses ranging between 32.2 and 41.6~$M_{\odot}$.
The grey region represents the observational constraints shown in Table~\ref{tab:observed_properties}.
}
\label{fig:logL_mass_postCE} 
\end{figure}

To model the post-CE phase of IC 10 X-1, we start by modeling the observed WR star.
We construct He star models with different initial masses at the measured metallicity for IC 10 (Z = 0.004, see Table~\ref{tab:observed_properties}) using the stellar evolutionary code
described in \S\ref{sec:methodology}. 
Then, we evolve the models and retain those that at a certain time during their evolution have a luminosity and mass in agreement with the observations listed in
Table~\ref{tab:observed_properties} (see Figure~\ref{fig:logL_mass_postCE}). Hereafter, we refer to these He star models as ``successful''.
At this stage, we use the observational constraint on the WR mass rather than effective temperature.
This is because the effective temperature shown in Table~\ref{tab:observed_properties} is indeed the spectroscopic temperature,
which is generally unreliable for WR stars \citep[see][]{crowther07}.

Next, we use the properties of the successful He star models to follow the evolution of the post-CE binary orbit and the He star's spin until the present time. Our goal is to constrain the
post-CE binary properties.
For each successful He star model, we vary the post-CE orbital semi-major axis in steps of 0.1~$R_\sun$ and consider different BH masses according to the measured mass
function ($f(M_{BH})$) and inclination angle ($i$) listed in Table~\ref{tab:observed_properties}. Specifically, we take the mass of each successful He star model at the current epoch and
compute the maximum and minimum BH mass at present ($M_{BH, max}$ and $M_{BH, min}$, respectively) using the measured $f(M_{BH})$ and $i$ with $1\sigma$ of observational
uncertainties. Then, we define
\begin{align}
M_{BH, mean} &\equiv \frac{M_{BH, max}+ M_{BH, min}}{2},\\
\Delta M_{BH} &\equiv \frac{M_{BH, max} - M_{BH, mean}}{2},
\end{align}
and vary the BH mass at present from $M_{BH, min}$ to $M_{BH, max}$, in steps of $0.2 \cdot \Delta M_{\rm BH}$.
The amount of mass accreted onto the BH since its birth is negligible, as the \cite{BondiHoyle1944} accretion of the stellar-wind material leads to a very small capturing fraction.
Hence, the post-CE BH mass is only slightly different from the one at present.
For the post-CE configuration of the binary orbit and the He star's spin, we assume that CE evolution circularizes and synchronizes the binary orbit.
We note that our assumption that the post-CE He star's spin is synchronized with the orbital frequency is completely arbitrary. However, this arbitrary choice only affects the strength
of the tide exerted on the He star by the BH.
This tide is weak in general, as the orbit of the BH--He star binary in this study is not very tight.
From post-CE to current epoch, the He star is always far from filling its Roche lobe, with the Roche lobe radius being roughly 3 to 9 times larger than the He star's radius. 
Hence, our arbitrary assumption of the post-CE He star's spin does not have any significant influence on our analysis.
Under these assumptions, we follow the evolution of the binary orbital semi-major axis and eccentricity, and the He star's spin,
accounting for tides exerted on the He star by the BH, stellar wind mass loss, orbital angular momentum loss due to gravitational radiation, and accretion of the He star's wind material
onto the BH.
In order to determine how much wind material is accreted onto the BH, we compute the accretion rate $\dot{M}_{acc}$.
Specifically, we adopt the \cite{BondiHoyle1944} accretion model and follow the formalism in \S~4.1 and 4.2 of \cite{BelczynskiKRTABMI2008} to obtain $\dot{M}_{acc}$.

The relevant ordinary differential equations (ODEs) governing the orbital evolution are integrated forward in time. To do this, we use the orbital evolution code described in the supplementary
information of \cite{ValsecchiNature2010} with the following two modifications.
\begin{enumerate}
\item
For the second-order tidal coefficient $E_{2}$, we take a stellar model from \cite{Claret2005} with an initial mass of 25.2~$M_{\odot}$ and metallicity of 0.004 (both in agreement with the
observations shown in Table~\ref{tab:observed_properties}), and derive
\begin{align}
{\rm log}(E_{2}) =  &-5.49491 \nonumber \\
&- 1.94284 t_{MS}^{51.277} -1.99707 t_{MS}^{2.41139},
\end{align}
where $t_{MS}$ is the star's evolutionary time expressed in units of the main sequence lifetime. As we are dealing with a He star, $t_{\rm MS}$ is taken to be the time the star has spent
burning He at its center.

\item
For the effect of wind mass loss on the spin of the He star, we approximate that the wind carries away its angular momentum from a thin shell at the star's surface.
Hence, we set 
\begin{equation}
\dot{J}_{spin} =  \frac{2}{3}\dot{M_{2}}R_{2}^2\omega,
\end{equation}
where $J_{spin}$ is the spin angular momentum, $R_{2}$ is the stellar radius and $\omega$ is the spin angular frequency of the He star .
\end{enumerate}
For each combination of post-CE binary component masses and orbital semi-major axis, the integration of the relevant ODEs proceeds forward in time only if the He star is spinning
slower than the break-up frequency $\Omega_c \approx \sqrt{G M/R^3}$ and is not filling its Roche lobe.

\begin{figure}
\begin{center}
\epsscale{1.}
\plotone{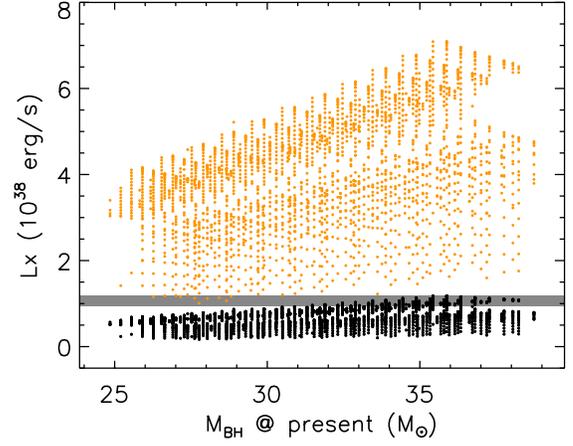}
\caption{
The variation of our predicted X-ray luminosity against the BH mass at present for every successful evolutionary sequences.
We consider two extreme values for the spin of the BH (see Equation~\ref{eqn:Lx}): non-spinning (black dots) and maximally spinning (orange dots). 
The grey region represents the observed X-ray luminosity presented in Table~\ref{tab:observed_properties}.}
\label{fig:Lx_vs_MBHnow} 
\end{center}
\end{figure}

\begin{figure}
\begin{center}
\epsscale{1.}
\plotone{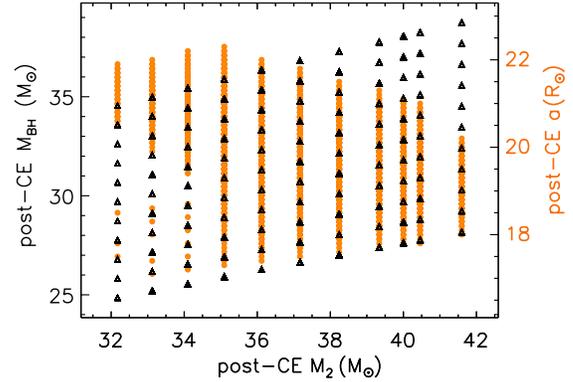}
\caption{
Post-CE binary component masses and semi-major axis of IC 10 X-1 given by all successful evolutionary sequences.
}
\label{fig:Masses_and_orbSet_postCE} 
\end{center}
\end{figure}

At the end of the calculation, we retain the evolutionary sequences that at the present time match the observed orbital period. Hereafter,
we refer to them as ``successful'' evolutionary sequences.
The X-ray luminosity at the present time of each evolutionary sequence is computed by following the formalism in \S~9.1 of \cite{BelczynskiKRTABMI2008}. Using the relation between
the radius of the accretor and the unknown BH spin, the X-ray luminosity is given by
\begin{equation}
L_{X} = \eta_{bol} \epsilon \frac{\dot{M}_{BH}c^{2}}{r_{isco}},
\label{eqn:Lx}
\end{equation}
where $c$ is the speed of light, and $\eta_{bol}$ and $\epsilon$ are adopted to be 0.8 and 0.5, respectively. The variable $r_{isco}$ is the radius of inner most stable circular orbit
around the BH expressed in units of $GM_{BH}/c^2$. It equals 6 for a non-spinning BH and 1 for a maximally spinning BH.
Figure~\ref{fig:Lx_vs_MBHnow} shows that for every successful evolutionary sequence the observed X-ray luminosity always falls in between the upper ($r_{isco} = 1$) and lower
($r_{isco} = 6$) limits of our predicted $L_X$ at the present time. Hence, we are not formally imposing the observed $L_X$ as a constraint, but our predicted $L_X$ at the present time
of all successful evolutionary sequences are naturally consistent with observed values.

The post-CE binary component masses and orbital semi-major axis of IC 10 X-1 given by all successful sequences are illustrated in
Figure~\ref{fig:Masses_and_orbSet_postCE}.
The figure shows that the post-CE binary consists of a 25--39~$M_\sun$ BH and a 32--42$~M_\sun$ He star and has  an orbital separation of 17--22~$R_\sun$.
The equivalent orbital period is 25--35~hr.
Due to the intense stellar wind suffered by the He star, the orbital separation increases continuously. Such a binary evolves to become the observed XRB
IC\;10\;X-1 in $\le 0.4$~Myr.

%%%%%%%%%%%%%%%%%%%%%
%%%  Beginning of CE Formalism   %%%
%%%%%%%%%%%%%%%%%%%%%
\section{Common Envelope Evolution Formalism}
\label{sec:cee_formalism}

Prior to the Monte Carlo simulation of the orbital dynamics involved in the core collapse event, we study whether our derived post-CE binary properties of IC 10 X-1 is achievable
based on the current understanding of CE evolution.
We construct stellar models of pre-CE He star progenitor candidates, which have core masses covering the entire range of post-CE He star's mass given by the successful
evolutionary sequences. These models are built based on the stellar evolution of isolated stars. We note that the BH progenitor could have potentially transferred mass to its 
companion before the core collapse event. However, we assume the He star progenitor will quickly adjust itself after the end of mass transfer. Hence, its structure right
before the CE phase is not expected to be significantly different from an isolated star of the same mass.
Furthermore, in order to initiate CE evolution, the immediate progenitor of the He star needs to fill its Roche lobe and start unstable mass transfer before its radius reaches its maximum value at
the instant $t_{Rmax}$. If this is not the case, the star will shrink rapidly afterward due to intense stellar wind mass loss. Meanwhile, most of this wind material will leave the binary
system, resulting in the widening of the binary orbit. Because of the increase in orbital separation, the Roche lobe of the star expands. Since the star is shrinking and its Roche lobe
is expanding, it cannot overfill its Roche lobe and initiate CE evolution after $t_{Rmax}$. Thus, we only consider the properties of our He star progenitor models before $t_{Rmax}$ when
studying the CE event.

\begin{figure}
\begin{center}
\plotone{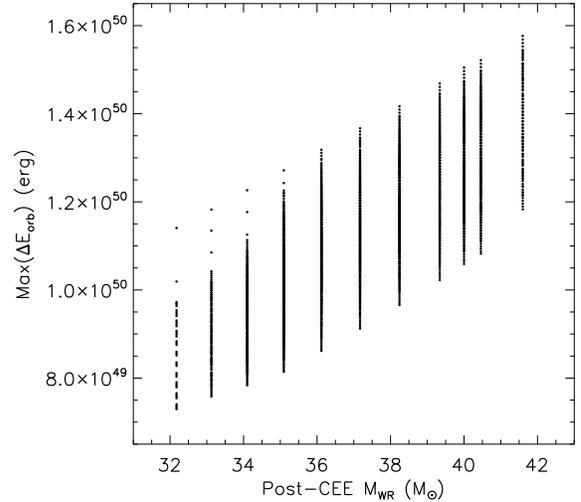}
\caption{Variation of maximum $\Delta E_{orb}$ against the mass of the post-CE He star ($M_{WR}$).
The maximum $\Delta E_{orb}$ is computed with Equation~(\ref{eqn:Eorb}), under the assumption that the pre-CE semi-major axis ($A_{preCE}$) approaches infinity.
The post-CE parameters are given by successful evolutionary sequences.}
\label{fig:fig1.eps}
\end{center}
\end{figure}

To examine whether CE evolution can explain our derived post-CE properties of IC 10 X-1,
we use the energy formalism \citep{webbink84} and compute the corresponding common envelope efficiency $\alpha_{CE}$,
\begin{equation}
\alpha_{CE} \cdot \Delta E_{orb} = E_{bind},
\label{eqn:ceerg}
\end{equation}
where $\Delta E_{orb}$ is the change in orbital energy and $E_{bind}$ is the energy required for dispersing the envelope to infinity. Here, $\Delta E_{orb}$ is simply
\begin{equation}
\Delta E_{orb} = -\frac{GM_{BH}M_{WRpro}}{2A_{preCE}} + \frac{GM_{BH}M_{WR}}{2A_{postCE}},
\label{eqn:Eorb}
\end{equation}
where $M_{WRpro}$ and $M_{WR}$ are the masses of the pre-CE He star progenitor and the post-CE He star, and $A_{preCE}$ and $A_{postCE}$ are the pre- and post-CE
orbital semi-major axis, respectively. Since the outcome of CE phase is given by the successful evolutionary sequences, the second term in Equation~(\ref{eqn:Eorb}) is predetermined.
Hence, it is obvious that $\Delta E_{orb}$ reaches maximum when $A_{preCE}$ approaches $\infty$.
Depending on the post-CE parameters given by each successful evolutionary sequence, maximum $\Delta E_{orb}$ ranges from $7.3 \times 10^{49}$ to $1.6 \times 10^{50}$ erg
(see Figure~\ref{fig:fig1.eps}).

When computing $E_{bind}$ of our He immediate progenitor models, we set the core-envelope boundary at the base of the convective thick hydrogen burning shell. Since this is
generally located close to where the hydrogen abundance ($X_H$) is $10\%$, we simply define the core boundary at $X_H = 0.1$.
\cite{dewi-tauris00} also considered the same definition of core boundary in their studies of CE evolution.
Our choice of core boundary definition is justified in \S\,\ref{sec:CE_trial2}.

Based on our adopted energy formalism shown as Equation~(\ref{eqn:ceerg}),
$\alpha_{CE} \le 1$ means the loss in the orbital energy is sufficient to unbind to envelope of our He star immediate progenitor model.
In order words, this means CE evolution is capable of producing our derived post-CE properties of IC 10 X-1.
On the other hand, $\alpha_{CE} > 1$ means that there is not enough energy to unbind the envelope and the pre-CE binary will eventually merge instead.
We examine five different treatments of CE evolution: the original \cite{webbink84} prescription (\S\ref{sec:CE_trial1}),
the \cite{webbink84} prescription with enhanced convective overshooting (\S\ref{sec:CE_trial2}) or increased mass-loss rates (\S\ref{sec:CE_massloss}),
allowing hyper-critical accretion onto the BH during CE evolution (\S\ref{sec:CE_trial3}),
and the ``enthalpy'' formalism \citep{ivanova-chaichenets11} (\S\ref{sec:CE_trial4}).
We find that only the ``enthalpy'' formalism is capable of explaining the post-CE binary properties given by all successful evolutionary sequences with $\alpha_{CE} \le 1$.
In the following subsections, we will discuss each of the aforementioned treatments of CE evolution.

\begin{deluxetable*}{ccccccccc}
\tablewidth{18 cm}
\tablecolumns{8}
\tablecaption{Properties of Selected He Star Progenitor Models}
\tablehead{
\colhead{Model} & \colhead{Initial Mass ($M_\sun$)} & \colhead{$\alpha_{ov}$\tablenotemark{$a$}} &
\colhead{$M_{WRpro}$ ($M_\sun$)\tablenotemark{$b$}} & \colhead{$M_{core}$ ($M_\sun$)\tablenotemark{$c$}} & \colhead{$R_{WRpro}$ ($R_\sun$)\tablenotemark{$d$}} &
\colhead{$E_{bind}$ ($10^{50}$ erg)\tablenotemark{$e$}} & \colhead{max. $\Delta E_{orb}$ ($10^{50}$ erg)\tablenotemark{$f$}}
}
\startdata
I & 80.4 & 0.12 & 65.9 & 32.2 & 1406 & 4.05 & 1.14\\
II & 84.1 & 0.12 & 67.3 & 34.1 & 1337 & 4.33 & 1.23\\
III & 88.6 & 0.12 & 69.9 & 36.1 & 1268 & 4.72 & 1.32\\
IV & 92.6 & 0.12 & 72.9 & 38.2 & 1214 & 5.13 & 1.42\\
V & 95.9 & 0.12 & 75.6 & 40.0 & 1175 & 5.48 & 1.50\\
VI & 99.9 & 0.12 & 78.2 & 41.6 & 1136 & 5.85 & 1.58\\
\tableline
VII & 75.0 & 0.20 & 60.0 & 32.2 & 1429 & 3.84 & 1.14\\
VIII & 78.5 & 0.20 & 62.4 & 34.1 & 1343 & 4.14 & 1.23\\
IX & 83.0 & 0.20 & 66.3 & 36.1 & 1262 & 4.63 & 1.32\\
X & 86.8 & 0.20 & 68.2 & 38.2 & 1205 & 4.85 & 1.42\\
XI & 89.9 & 0.20 & 70.9 & 40.0 & 1171 & 5.16 & 1.50\\
XII & 93.6 & 0.20 & 73.0 & 41.6 & 1140 & 5.46 & 1.58\\
\tableline
XIII & 67.0 & 0.30 & 53.8 & 32.2 & 1403 & 3.49 & 1.14\\
XIV & 70.9 & 0.30 & 56.9 & 34.1 & 1317 & 3.82 & 1.23\\
XV & 74.8 & 0.30 & 59.7 & 36.1 & 1239 & 4.13 & 1.32\\
XVI & 78.6 & 0.30 & 62.4 & 38.2 & 1188 & 4.40 & 1.42\\
XVII & 81.8 & 0.30 & 65.1 & 40.0 & 1149 & 4.69 & 1.50\\
XVIII & 84.8 & 0.30 & 67.0 & 41.6 & 1118 & 4.92 & 1.58
\enddata
\label{tab:cetest}
\tablecomments{The models with $\alpha_{ov} = 0.12$ have normal strength of convective core overshooting.
The parameters listed in column 4--7 are the properties of the He star progenitor models when their radii reach their maximum values at the instant $t_{Rmax}$.}
\tablenotetext{$a$}{convective overshooting parameter}
\tablenotetext{$b$}{mass of the He star immediate progenitor}
\tablenotetext{$c$}{core mass of the He star immediate progenitor (same as the mass of the descendent He star right after CE evolution)}
\tablenotetext{$d$}{radius of the He star immediate progenitor}
\tablenotetext{$e$}{energy required to disperse envelope of the He star immediate progenitor}
\tablenotetext{$f$}{maximum change in orbital energy during CE evolution involving the He star progenitor at $t_{Rmax}$ (see Figure~\ref{fig:fig1.eps})}
\end{deluxetable*}

\subsection{Original \cite{webbink84} Prescription}
\label{sec:CE_trial1}

\begin{figure}
\begin{center}
\plotone{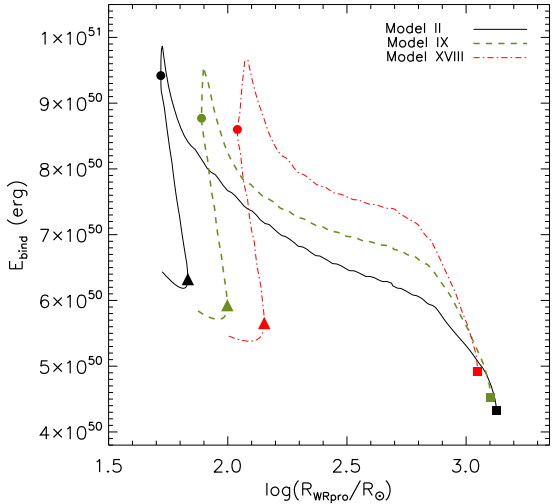}
\caption{
The variation of $E_{bind}$ against log($R_{WRpro}$) throughout the evolution of three different He star progenitor models (Models II, IX, XVIII as listed in
Table~\ref{tab:cetest}), which have similar initial masses but different strength of convective overshooting.
Different events in stellar evolution are illustrated on each curve: the termination of main-sequence (triangles), the beginning of thick H shell burning (circles) and the time of their stellar
radii reaching maximum(squares).
Throughout the evolution of these models, their $E_{bind}$ have similar behavior and reach the minimum when the stellar radii of these models reach maximum.
}
\label{fig:fig2.eps}
\end{center}
\end{figure}

First, we use the standard definition of $E_{bind}$, which can be written as
\begin{equation}
E_{bind} = -\int^{surface}_{core\;boundary} \bigl (\Phi(m) + \epsilon(m) \bigr )\; dm.
\label{eqn:Ebind}
\end{equation}
Here $\Phi(m) = -Gm/r$ is the gravitational potential and $\epsilon(m)$ is the specific internal energy,
which includes the thermal energy of the plasma gas only and does not include the recombination energy of H and He nor the association energy of $\rm H_2$ \citep{han...94, han...95}.
For our progenitor models, the sum of the recombination and association energies is less than a thousandth of the thermal energy of the plasma gas.
Hence, $E_{bind}$ will not change significantly by whether including these energy sources or not. 
In Figure~\ref{fig:fig2.eps}, we illustrate the typical variation of $E_{bind}$ throughout the evolution of a He star progenitor. After the termination of main-sequence at $t_{tms}$,
$E_{bind}$ increases as the He star progenitor shrinks. It reaches a maximum shortly after the beginning of convective thick H shell burning at $t_{Hsb}$. Then, it starts to decrease
because of the envelope expansion. Figure~\ref{fig:fig2.eps} also shows that throughout its evolution prior to $t_{Rmax}$, $E_{bind}$ of a He star progenitor is always the smallest
at $t_{Rmax}$.

Using the above definition of $E_{bind}$ and Equation~(\ref{eqn:ceerg}), we search for He star progenitors models that will give us $\alpha_{CE} \le 1$ at a certain time during its
evolution before $t_{Rmax}$. To achieve this goal, we create a grid of stellar models at the observed metallicity of IC 10 X-1 by varying the initial mass from 65 to 105~$M_\sun$
in steps of 0.1~$M_\sun$. This grid covers the entire initial mass range of massive stars, which can have core masses fall within the mass range of post-CE He stars constrained
by all successful evolutionary sequences (i.e. 32--42~$M_\sun$).

For each model in our grid, we match its core mass at any time before $t_{Rmax}$ with the post-CE He star masses of the successful evolutionary sequences. Then, we compute
$\Delta E_{orb}$ with Equation~(\ref{eqn:Eorb}) using the relevant post-CE binary properties given by those matched sequences. Using this $\Delta E_{orb}$ and the $E_{bind}$
defined by Equation~(\ref{eqn:Ebind}), we can obtain $\alpha_{CE}$ with Equation~(\ref{eqn:ceerg}). We find that for all models $\alpha_{CE} > 1$, as $E_{bind}$ is always several
times larger than $\Delta E_{orb}$. To illustrate this, we select six representative models of He star progenitors from our grid and list them in Table~\ref{tab:cetest} as Model I--VI.
Here, we consider the CE evolution involving these models at $t_{Rmax}$, which is the time when these models have the lowest $E_{bind}$. The mass range of the cores in these models
at $t_{Rmax}$ is the same as that of the post-CE He stars given by all successful evolutionary sequences. In column 8, we list the maximum $\Delta E_{orb}$ during the CE evolution
involving these models, under the assumption that $A_{preCE}$ approaches $\infty$ (see Figure~\ref{fig:fig1.eps}). It is obvious that for these models $E_{bind}$ is at least 3.5 times
larger than the corresponding maximum $\Delta E_{orb}$. Thus, $\alpha_{CE}$ is always $> 1$.

\subsection{Enhanced Convective Overshooting}
\label{sec:CE_trial2}

\begin{figure}
\begin{center}
\plotone{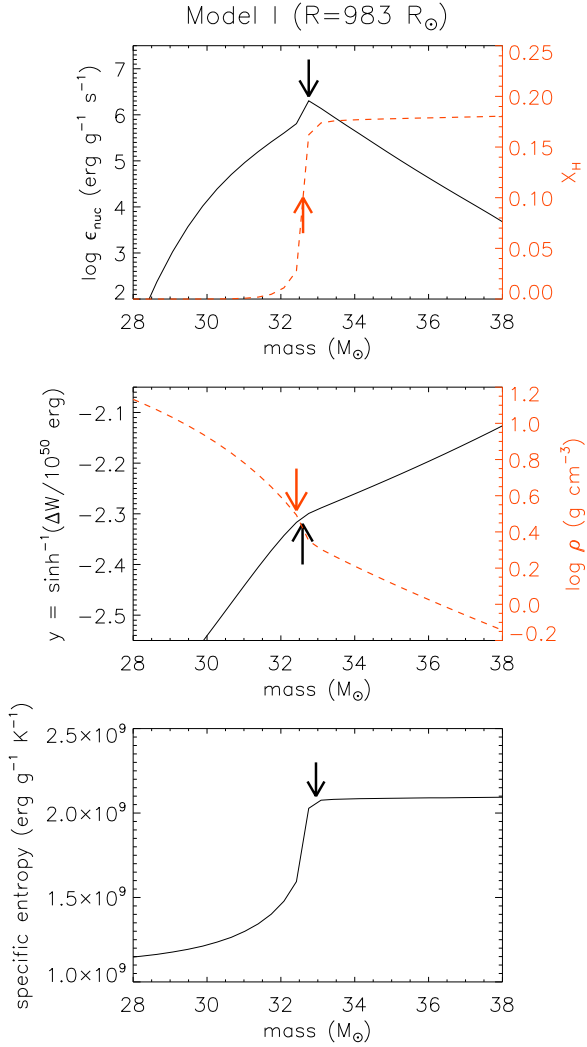}
\caption{
The internal structure of Model I (see Table\,\ref{tab:cetest}) when its radius is 983\,$R_\odot$.
The arrows on each panel indicate the core boundary given by different definitions (see text in \S\,\ref{sec:CE_trial2}),
showing that the core boundary does not vary significantly with different definitions.}
\label{fig:core_bndry.eps}
\end{center}
\end{figure}

A possible reason for the negative result in the previous trial could be that the envelopes of our He star progenitor models are too massive.
\cite{tauris-dewi01} showed that the envelope mass and $E_{bind}$ vary significantly with different definitions of core boundary.
To check whether there is a definition of core boundary that can decrease the $E_{bind}$ of our He star progenitor models, we follow \cite{tauris-dewi01} and consider these
definitions of core boundary: energy production rate (max $\epsilon_{nuc}$), binding energy profile \citep{han...94},
mass-density gradient \citep[$\partial^2 \log \rho / \partial m^2 = 0$,][]{bisscheroux98}, and specific entropy profile.
We find that the $E_{bind}$ of our He-star progenitor models resulted from these core boundary definitions are similar to those obtained from our canonical choice
($X_H$ = 0.1), with a difference of $<\,3\%$.
In Figure\,\ref{fig:core_bndry.eps}, we use Model I listed in Table\,\ref{tab:cetest} as an example to illustrate how similar the locations of the core boundary given by different definitions are.

Alternatively, we can reduce the envelope mass by increasing the strength of convective core overshooting.
In this trial, we increase the convective core overshooting ($\alpha_{ov}$) parameter from the canonical value of 0.12 to 0.2 and 0.3.
Our canonical $\alpha_{ov} = 0.12$ leads to an overshooting length of $0.32\,H_p$, where $H_p$ is the pressure scale height \citep[see][]{schroder...97}.
We again compute $\alpha_{CE}$ using Equation~(\ref{eqn:ceerg}) with $E_{bind}$ defining as Equation~(\ref{eqn:Ebind}).

As before, we make grids of He star progenitor models at the observed metallicity of IC 10 X-1 with $\alpha_{ov} = 0.2$ and 0.3, by varying the initial mass
between 60--95 and 55--88~$M_\sun$ in steps of 0.1~$M_\sun$, respectively. Each grid covers the entire initial mass range of massive stars, which can have core masses fall
within the mass range of the post-CE He stars constrained by all successful evolutionary sequences (i.e. 32--42~$M_\sun$).

After computing $\alpha_{CE}$ in the same way as our previous trial (see \S\ref{sec:CE_trial1}) for all He star progenitor models in our grids of $\alpha_{ov} = 0.2$ or 0.3,
we do not find any cases of $\alpha_{CE} \le 1$. To help explaining this result, we select a group of representative models from each grid and list them as Model VII--XII ($\alpha_{ov} = 0.2$)
and Model XIII--XVIII ($\alpha_{ov} = 0.3$) in Table~\ref{tab:cetest}. Here, we consider the CE evolution involving these models at $t_{Rmax}$, which is the time when these models
have the lowest $E_{bind}$ (see Figure~\ref{fig:fig2.eps}). For each group of models, their core masses cover the entire range of post-CE He star's mass given by the successful
evolutionary sequences. The maximum $\Delta E_{orb}$ during the CE evolution involving these models (see Figure~\ref{fig:fig1.eps}) are listed in column 8. When comparing the $E_{bind}$
of these models to that of the models with the normal strength of convective overshooting (i.e. Model I--VI), we find that for the models with the same core mass $E_{bind}$ only
decreases slightly with the increasing $\alpha_{ov}$. Considering the difference between $E_{bind}$ and maximum $\Delta E_{orb}$ of these models, that amount of decrease in
$E_{bind}$ is not enough to make $\alpha_{CE} \le 1$. 

\subsection{Enhanced Mass Loss Rates}
\label{sec:CE_massloss}

As we have mentioned in \S\,\ref{sec:methodology}, the mass loss rates for massive stars are not well constrained.
The uncertainty in these rates not only can affect the envelope mass of the He star progenitor that needs to be ejected in CE evolution,
but also the density structure of the envelope and hence $E_{bind}$ \citep[see e.g.][]{podsiadlowski...03}.
To study the variance in $E_{bind}$ due to this uncertainty, we compute He star progenitor models with all considered mass loss rates being enhanced by a factor of two.
When comparing these models to our canonical models ($\alpha_{ov} = 0.12$) according to the core mass at $t_{Rmax}$, we find that the $E_{bind}$ of these models are
3--14\% smaller.
However, this decrease in $E_{bind}$ is not sufficient to make $\alpha_{CE} \le 1$.

On the other hand, there is also uncertainty in the mass loss rate for WR stars.
If it is higher than what we adopted when modeling the post-CE binary (see \S\,\ref{sec:postCE_binary_modeling}), the post-CE orbit can be tighter than what we found.
This is because the post-CE orbital evolution is dominated by wind mass loss from the system, which always increases the orbital separation.
To get a flavor on how much this uncertain can change the post-CE orbital separation, and hence the available orbital energy for ejecting the envelope in CE evolution,
we let the mass loss rate for WR stars to be a factor of two higher than what we consider.
Then, the new mass of the post-CE He star ($M'_{He,postCE}$) can be written as
\begin{equation}
M'_{WR,postCE} = 2M_{WR,postCE} - M_{WR,now},
\label{eqn:sw1}
\end{equation}
where $M_{WR,now}$ is the observed mass of the He star in IC\,10\,X-1 and $M_{WR,postCE}$ is the mass of the post-CE He star derived from our post-CE binary modeling.
Using the equation of Jeans-mode mass loss \citep[see e.g.][]{BelczynskiKRTABMI2008}, we can write the new post-CE orbital separation ($A'_{postCE}$) as
\begin{equation}
\frac{A'_{postCE}}{A_{postCE}} = \frac{M_{BH} + M_{WR,postCE}}{M_{BH}+M'_{WR,postCE}}.
\label{eqn:sw2}
\end{equation}
With Equation\,(\ref{eqn:Eorb}) and taking the limit that $A_{preCE}$ tends to $\infty$, we can express the new maximum change of orbital energy during CE evolution
(max. $\Delta E'_{orb}$) as
\begin{equation}
max.\;\Delta E'_{orb} = \frac{M'_{WR,postCE}}{M_{WR,postCE}} \frac{A_{postCE}}{A'_{postCE}} \cdot max.\;\Delta E_{orb}
\label{eqn:sw3}
\end{equation}
Using Equation\,(\ref{eqn:sw1})--(\ref{eqn:sw3}), we find that doubling our adopted mass loss rates for WR stars will lead to <\,45\% increase in the maximum change
of orbital energy during CE evolution.
This increase is not enough to make $\alpha_{CE} \le 1$, as $E_{bind}$ of our He star progenitor models are at least 3 times larger than the original max. $\Delta E_{orb}$
(see Table\,\ref{tab:cetest}).

We show that the required $\alpha_{CE}$ will still be $> 1$ even if we increase our adopted mass loss rates by a factor of 2.
However, we note that the uncertainties in these mass loss rates are larger than what we consider, especially for the stars evolved off the main sequence.
For instance, when studying the mass transfer in massive binaries, \cite{petrovic...05} allowed a factor of 6 uncertainty in their mass loss rates for WR stars.
Given these large uncertainties, we can potentially obtain $\alpha_{CE} < 1$ by adopting mass loss rates at even higher values.
Since it is numerically challenging to evolve very massive stars with extraordinarily high mass loss rates beyond the main sequence,
we choose to seek an alternative CE treatment that will naturally give $\alpha_{CE} < 1$.

\subsection{Hyper-Critical Accretion}
\label{sec:CE_trial3}

It has been suggested that a compact object might accrete a significant amount of mass after being engulfed into the envelope of its companion, due to hyper-critical accretion
\citep{blondin86, chevalier89, chevalier93, brown95}. Since only part of the companion's envelope will be dispersed to infinity, the $E_{bind}$ in the energy formalism shown as
Equation~(\ref{eqn:ceerg}) needs to be adjusted accordingly. Hence, we write the equation of energy balance as
\begin{equation}
\alpha_{CE} \cdot \Delta E_{orb} = \Delta E_{bind} = f_{ej} \cdot E_{bind},
\label{eqn:mod_ceerg}
\end{equation}
where
\begin{equation}
f_{ej} \equiv \frac{M_{env}  - \Delta M_{BH}}{M_{env}}
\label{eqn:ejectedMF}
\end{equation}
is the fraction of envelope mass ejected to infinity. Here, $M_{env}$ is the mass of the envelope and $\Delta M_{BH}$ is the amount of mass accreted onto the BH.
Based on Equation~(\ref{eqn:mod_ceerg}) and (\ref{eqn:ejectedMF}), we follow \cite{belczynski...02} and derive the rates of change in the BH mass and the binary semi-major axis 
during the phase of hyper-critical accretion.
The detailed derivation of these rates with respect to the mass of the He star progenitor ($M_{com}$) during the accretion phase can be found in the Appendix.
Our rates are different from those derived by \cite{belczynski...02}, since we consider a fraction instead of the total $E_{bind}$ when balancing the energy budget of envelope ejection.
In other words, we use Equation~(\ref{eqn:mod_ceerg}) instead of Equation~(\ref{eqn:ceerg}) to incorporate $\alpha_{CE}$.

Using the He star progenitor models in our constructed grids (see \S\ref{sec:CE_trial1} and \ref{sec:CE_trial2}) and considering their properties at any time prior to
$t_{Rmax}$, we numerically integrate Equation~(\ref{eqn:dA_dM}) and (\ref{eqn:dMbh_dM}) from $M_{com} = M_{core}$ to $M_{WRpro}$ with different $\alpha_{CE}$ between 0 and 1.
Here, we assume that $\alpha_{CE}$ is constant throughout the whole CE evolution, including the hyper-critical accretion phase.
From this integration, we can obtain the required radius of the He star progenitor and the BH mass at the onset of the hyper-critical accretion phase (i.e. when the BH touches its surface).
Then, using Equation\,(\ref{eqn:hyp_aprece}), we can compute the corresponding pre-CE binary semi-major axis and justify whether it is physically possible. 

We find that the BH could have accreted $\sim10\,M_\sun$ during the hyper-critical accretion phase.
However, the radius of the He star progenitor at the onset of the hyper-critical accretion phase is required to be at least 1.3 times larger than that at the beginning of CE evolution
(i.e. when the He star progenitor fills its Roche lobe).
According to Equation~(\ref{eqn:hyp_aprece}), there does not exist a pre-CE binary semi-major axis leading to such condition.
Therefore, we conclude that the hyper-critical accretion formalism cannot explain the post-CE binary properties given by the successful evolutionary sequences.

\subsection{``Enthalpy'' Formalism}
\label{sec:CE_trial4}

\begin{figure}
\begin{center}
\plotone{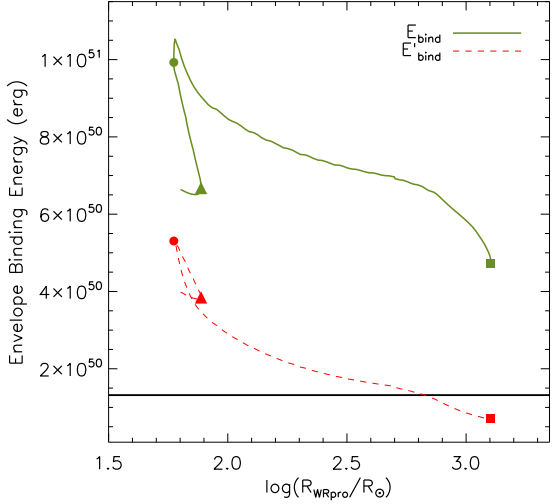}
\caption{
The variation of $E_{bind}$ and $E'_{bind}$ (defined by Eqn\;(\ref{eqn:Ebind}) and (\ref{eqn:E'bind}), respectively) against log($R_{WRpro}$) throughout the evolution of Model III.
Different events in stellar evolution are illustrated on each curve: the termination of main-sequence (triangles), the beginning of convective thick H shell burning (circles) and the time
of its stellar radius reaching maximum (squares).
The horizontal line indicates the maximum $\Delta E_{orb}$ during CE evolution involving this He star progenitor model (see text in \S\ref{sec:CE_trial1}).
According to Equation~(\ref{eqn:ceerg}), this model will give us $\alpha_{CE} < 1$ at $t_{Rmax}$ using $E'_{bind}$ as the description of envelope binding energy.
}
\label{fig:fig5.eps}
\end{center}
\end{figure}

\cite{ivanova-chaichenets11} argued that enthalpy should be considered when calculating the binding energy of the envelope and introduced
\begin{equation}
E'_{bind} = -\int^{surface}_{core\;boundary} \left (\Phi(m) + \epsilon(m) + \frac{P(m)}{\rho (m)} \right)\; dm.
\label{eqn:E'bind}
\end{equation}
Using Model III listed in Table~\ref{tab:cetest} as an example, we illustrate the behavior of its envelope binding energy computed by Equation~(\ref{eqn:Ebind}) and (\ref{eqn:E'bind})
on Figure~\ref{fig:fig5.eps}. It is clear that $E'_{bind}$ is always smaller than $E_{bind}$ at any time during the evolution of Model III. This is because the term $P(m)/ \rho (m)$ is always
positive everywhere within a star. Including it as an energy source lowers the binding energy of the envelope, and hence the required CE efficiency.
Figure~\ref{fig:fig5.eps} also shows that $E'_{bind}$ at $t_{Rmax}$ is smaller than the corresponding maximum $\Delta E_{orb}$ during the CE phase (see Figure~\ref{fig:fig1.eps}
and Table~\ref{tab:cetest}).
According to Equation~(\ref{eqn:ceerg}), $\alpha_{CE}$ could be $\le 1$ if $\Delta E_{orb}$ during the CE evolution involving this progenitor model is close to that maximum value.

To examine whether the ``enthalpy'' formalism can explain our derived post-CE properties of IC 10 X-1, we consider the models of He star progenitors in our constructed grid with
the normal strength of convective overshooting ($\alpha_{ov} = 0.12$, see \S\ref{sec:CE_trial1}). Throughout the evolution of each model, we match its core mass with the masses
of the post-CE He stars given by the successful evolutionary sequences. Then we compute $\alpha_{CE}$ with Equation\;(\ref{eqn:ceerg}) and (\ref{eqn:E'bind}), using the post-CE
binary properties of the matched sequences. As we expected, we find many cases with $\alpha_{CE} \le 1$.

To summarize, the post-CE binary properties given by the successful evolutionary sequences can be explained by our understanding of CE evolution, using the standard $\alpha_{CE}$
prescription \citep{webbink84} with the ``enthalpy'' formalism \citep{ivanova-chaichenets11}.
%The ``enthalpy'' formalism appears to be an appropriate description on the energy budget of envelope ejection during CE evolution,
%at least for the massive stars involved in the formation of IC 10 X-1.
Since the ``enthalpy'' formalism can account for the formation of IC 10 X-1, we suggest that this strengthens the evidence in favor of the ``enthalpy'' formalism being an appropriate
description of the energy budget of envelope ejection during CE evolution, at least for the massive stars involved in the formation of IC 10 X-1.
We will discuss alternative CE treatments in \S\;\ref{sec:conclusion}, which are not considered in this study but might also be able to explain the existence of IC\;10\;X-1 without
the need to invoke $\alpha_{CE} > 1$.
In addition, we find that the core boundary of our He star progenitor models are roughly unchanged when considering alternative boundary definitions existed in the literature.

%%%%%%%%%%%%%%%%%%%%%%%%%%%%%%
%%%   Beginning of orbital dynamics at core collapse  %%%
%%%%%%%%%%%%%%%%%%%%%%%%%%%%%%
\section{Orbital Dynamics at Core Collapse}
\label{sec:dynamics@CC}

Using the progenitor properties obtained from the previous steps (see \S\ref{sec:postCE_binary_modeling} and \ref{sec:cee_formalism}) as constraints,
we perform Monte Carlo simulations on the binary orbital dynamics involved in the core collapse event.
Our goal is to derive constraints on the properties of the BH immediate progenitor and the magnitude of the natal kick imparted to the BH.

Just before the core collapse event, the binary consists of the BH immediate progenitor and its companion star, with an orbital semi-major axis $A_{preSN}$ and
eccentricity $e_{preSN}$.
Note that since it is not necessary that the pre-SN progenitor has experienced any mass-transfer phase, we cannot assume that the pre-SN orbit is circular.
Instead we consider the full range of possibilities with eccentric pre-SN orbits.
The masses of the BH immediate progenitor and its companion are $M_{BHpro}$ and $M_2$, respectively.
During the core collapse event, the mass loss from the BH immediate progenitor and the potential natal kick imparted to the BH alter the binary orbital properties.
Hence, the post-SN orbital semi-major axis and eccentricity become $A_{postSN}$ and $e_{postSN}$, respectively. As we assume the companion star is not affected by the
instantaneous core collapse event, its properties remain unchanged.

Using the equation of binary orbital energy and angular momentum, the pre- and post-SN binary properties are related as \citep{hills83, wong...12}:
\begin{align}
V^2_k  +  V^2_{BHpro}  +  2 V_k & V_{BHpro} \cos \theta_k \nonumber \\
	& = G(M_{BH}+M_2)\left (\frac{2}{r} - \frac{1}{A_{postSN}} \right )
\label{eqn:snEorb}
\end{align}
\begin{align}
G (M_{BH} + & M_2) A_{postSN}(1-e^2_{postSN}) \nonumber \\
	= r^2 \Bigl ( & V^2_k \sin^2 \theta_k \cos^2 \phi_k + \bigl [ \sin \psi (V_{BHpro} + V_k \cos \theta_k) \nonumber \\
	& - V_k \cos \psi \sin \theta_k \sin \phi_k \bigr ]^2 \Bigr ).
\label{eqn:snLorb}
\end{align}
Here, $V_k$ is the magnitude of the natal kick imparted to the BH, while $\theta_k$ and $\phi_k$ describe its direction in the frame of the BH immediate progenitor.
Specifically, $\theta_k$ is the polar angle of the natal kick with respect to the relative orbital velocity of the BH immediate progenitor,
and $\phi_k$ is the corresponding azimuthal angle \citep[see Figure 1 in][for a graphic representation]{kalogera00}.
The variable $r$, which is the separation between the BH immediate progenitor and its companion at the moment of core collapse, can be expressed as
\begin{equation}
r = A_{preSN} (1-e_{preSN}\cos \mathcal{E}_{preSN}),
\end{equation}
where the pre-SN eccentric anomaly $\mathcal{E}_{preSN}$ is related to the pre-SN mean anomaly $\mathcal{M}_{preSN}$ as
\begin{equation}
\mathcal{M}_{preSN} = \mathcal{E}_{preSN} - e_{preSN} \sin \mathcal{E}_{preSN}.
\end{equation}
The relative orbital speed $V_{BHpro}$ of the BH immediate progenitor in the pre-SN binary can be written as 
\begin{equation}
V_{BHpro} = \left [ G(M_{BHpro}+M_2) \Bigl ( \frac{2}{r} - \frac{1}{A_{preSN}}\Bigr ) \right ]^{1/2}
\end{equation}
Finally, the angle $\psi$ is the polar angle of position vector of the BH immediate progenitor with respect to its relative orbital velocity in its companion's frame,
which is related to the pre-SN orbital parameters as
\begin{equation}
\sin \psi = \left [ \frac{A^2_{preSN} (1-e^2_{preSN})}{r (2A_{preSN} - r)} \right ] ^{1/2}
\end{equation}

We start our calculation at the instant just before the core collapse event. The properties of the companion star are taken from a stellar model in a grid constructed by varying the
initial mass between 65--105 $M_\sun$, in steps of 0.1 $M_\sun$. It is indeed the same grid of models used in finding the correct formalism describing CE evolution, which is capable of
explaining the post-CE binary properties given by all successful evolutionary sequences (see \S\ref{sec:CE_trial1} and \ref{sec:CE_trial4}).
To obtain the properties of these companion star models at the core collapse event of the BH progenitor, we need to approximate when the BH is formed in the evolutionary time frame of
its companion ($t_{BH}$). Under the assumption that the BH progenitor and its companion are born at the same time, $t_{BH}$ simply equals to the lifetime of the BH progenitor.
We adopt that to be the lifetime of a stellar model with an initial mass of 150 $M_\sun$, which is approximately 2.9 Myr.
Other pre-SN binary properties, namely $M_{BHpro}$, $A_{preSN}$, $e_{preSN}$, $\mathcal{M}_{preSN}$, are drawn randomly from uniform distributions.
Supplemented with a natal kick magnitude $V_k$ and direction angles ($\theta_k$, $\phi_k$) drawn randomly from uniform and isotropic distributions, we can obtain $A_{postSN}$
and $e_{postSN}$ from the pre-SN binary properties using Equation~(\ref{eqn:snEorb}) and (\ref{eqn:snLorb}). 
In this calculation, $M_{BH}$ is taken directly from a successful evolutionary sequence, because the BH in our adopted formation scenario of IC 10 X-1 has only accreted negligible
amount of mass since its birth.

For each combination of companion star model and successful evolutionary sequence, we perform 2000 Monte Carlo trials. We only retain the data points that satisfy \emph{all} of the
following constraints and classify those data points as ``successful''.
\begin{enumerate}
\item
$M_{BHpro}$ is set to be less than 60 $M_\sun$, which is a conservative upper limit guided by the study of \cite{belczynski...10} on the maximum mass of stellar black holes.
We will discuss the impact of this limit on our derived constraints related to the BH formation in \S\ref{sec:CC_constraints}.
\item
The binary must survive through the core collapse event. This means $A_{postSN}$ and $e_{postSN}$ obtained from Equation~(\ref{eqn:snEorb}) and (\ref{eqn:snLorb}) needs
to have realistic values: $A_{postSN} > 0$ and $0 \le e_{postSN} < 1$.
\item
Since the core collapse event of the BH immediate progenitor is instantaneous, the separation between the pre-SN binary components is the same as that of the post-SN binary.
This gives a constraint as
\begin{align}
 A_{preSN} (1-e&_{preSN}\cos \mathcal{E}_{preSN}) \nonumber \\
 	&= A_{postSN} (1-e_{postSN}\cos \mathcal{E}_{postSN}),
\end{align}
which has to be satisfied with a realistic post-SN eccentric anomaly: $|\cos (\mathcal{E}_{postSN}) | \le 1$.
\item
Both pre-SN binary components cannot spin faster than the breakup angular velocity $\Omega_c \approx \sqrt{G M/R^3}$. Here, we assume their spins are pseudo-synchronized to the
pre-SN orbital angular velocity. Since the BH immediate progenitor is expected to be He rich due to potential binary interaction or intense mass loss via stellar wind, we approximate
its radius using Equation (3) in \cite{fryer-kalogera97}.
\item
According to our adopted formation scenario of IC 10 X-1, both components of the pre-SN binary need to fit within their Roche lobes at periapsis.

\begin{figure*}
\begin{center}
\plotone{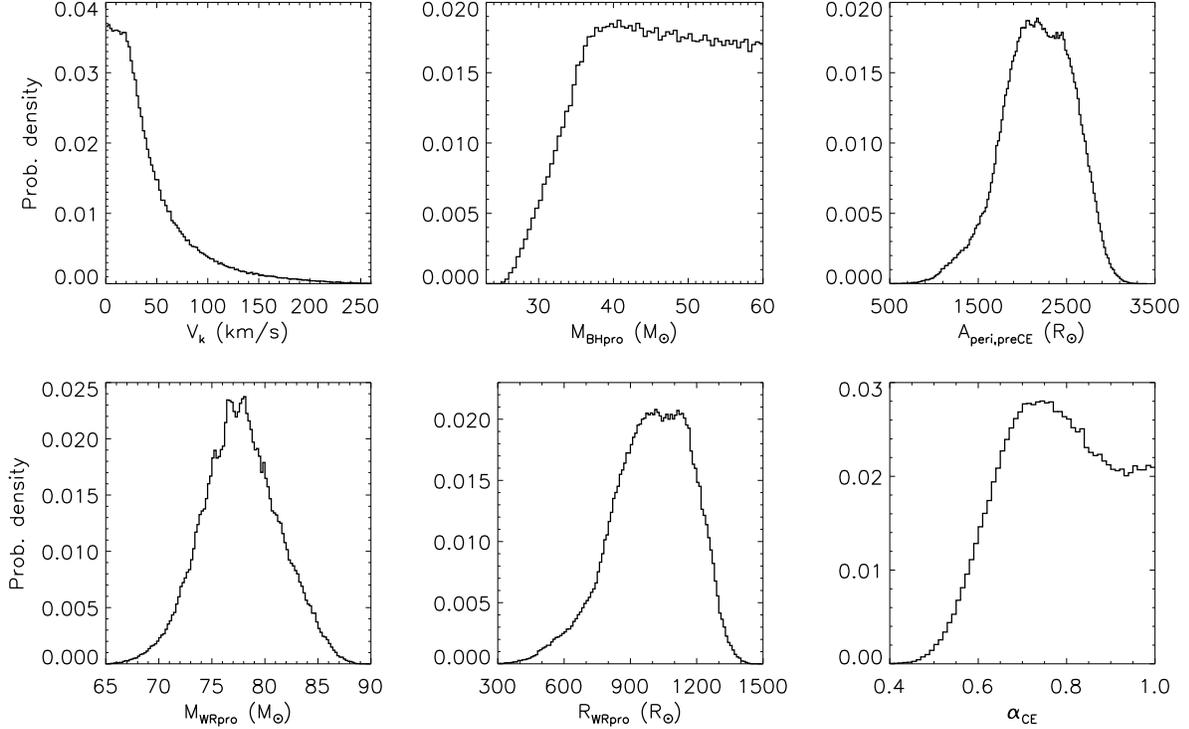}
\caption{Marginalized probability distribution functions (PDF) of different variables (from left to right):
(top row) the natal kick magnitude ($V_k$) imparted to the BH, the BH immediate progenitor mass ($M_{BHpro}$), the pre-CE orbital separation at periastron ($A_{peri,preCE}$), 
(bottom row) the pre-CE He star progenitor's mass ($M_{WRpro}$) and radius ($R_{WRpro}$), and the common envelope efficiency ($\alpha_{CE}$).}
\label{fig:1D-pdfs}
\end{center}
\end{figure*}

\item
The BH companion in the post-SN binary must later fill its Roche lobe at periapsis before its radius reaches its maximum value (see \S\ref{sec:cee_formalism}), which leads to CE evolution.
As massive stars evolve at roughly the same nuclear time scale, the time difference between the formation of the BH and the onset of CE event is small.
Hence, we assume the binary semi-major axis and eccentricity remain unchanged within this period of time.
The outcome of CE phase is indeed constrained by the post-CE binary properties of the corresponding successful evolutionary sequence.
When the BH companion fills its Roche lobe at periapsis, its core mass needs to match the corresponding mass of the post-CE He star, with a tolerance of 1~$M_\sun$.
This tolerance value is chosen according to the initial mass resolution in our grid of post-CE He star models, whose properties are used in constructing the successful evolutionary
sequences (see \S\ref{sec:postCE_binary_modeling}).
Furthermore, the common envelope efficiency $\alpha_{CE}$ determined by the standard $\alpha$ prescription \citep{webbink84} with the ``enthalpy'' formalism (see \S\ref{sec:CE_trial4})
must be $\le 1$.
\end{enumerate}

%%%%%%%%%%%%%%%%%%%%%
%%%   Beginning of Result Section  %%%
%%%%%%%%%%%%%%%%%%%%%
\section{Results}
\label{sec:results}

The elements presented in the previous sections can now be combined to establish a complete picture of how we track the evolution of IC 10 X-1 backwards in time and derive
constraints related to the BH formation in this system.
We first use the modeling of binary evolution and observational constraints to determine the post-CE binary properties. Specifically, our ``successful'' evolutionary sequences at the
present time simultaneously match the measured component masses, He star luminosity and binary orbital period of IC 10 X-1. Then, we search for the correct
formalism and treatment of CE evolution leading to the formation of IC 10 X-1. We find that the standard $\alpha$ prescription \citep{webbink84} with the ``enthalpy'' formalism
\citep{ivanova-chaichenets11} is capable of explaining the post-CE binary properties given by our successful evolutionary sequences. Last, we use our findings in the two
previous steps as part of the constraints applied on a Monte Carlo simulation of the binary orbital dynamics involved in the core collapse event. Each data point in this simulation
contains seven random parameters: the BH immediate progenitor mass, the pre-SN orbital mean anomaly, semi-major axis and eccentricity, the magnitude of the natal kick velocity
imparted to the BH and two angles describing the kick direction.
These random parameters are drawn from uniform or isotropic prior distributions.
If a data point satisfies all constraints mentioned in \S\ref{sec:dynamics@CC}, we classify it as a ``successful'' data point.
Our results and derived constraints (at 95.4\% of confidence) presented in what follows are obtained from all successful data points, as well as the marginalized PDFs illustrated in
Figure~\ref{fig:1D-pdfs}.

\subsection{Core Collapse Constraints}
\label{sec:CC_constraints}

Just before the core collapse event, we find the BH immediate progenitor mass ($M_{BHpro}$) to be $46 \pm 14~M_\sun$
and its highly evolved main-sequence companion mass ($M_2$) to be $82 \pm 7~M_\sun$.
At this time, the companion star has $\sim 4$--18\% of H left in its core.
The orbital separation ($r$) between the BH immediate progenitor and its companion is $8100_{-7100}^{+71000}~R_\sun$.
During the core collapse event, the BH immediate progenitor loses $\le 50 \%$ of its mass, which is $\le 20 \%$ of the total mass in the pre-SN binary.
Possible asymmetries developed in the core collapse event can lead to a natal kick ($V_k$) of $\le 130$~km/s imparted to the BH.
The PDFs of $V_k$ and $M_{BHpro}$ are presented in Figure~\ref{fig:1D-pdfs}.

\begin{figure}
\begin{center}
\plotone{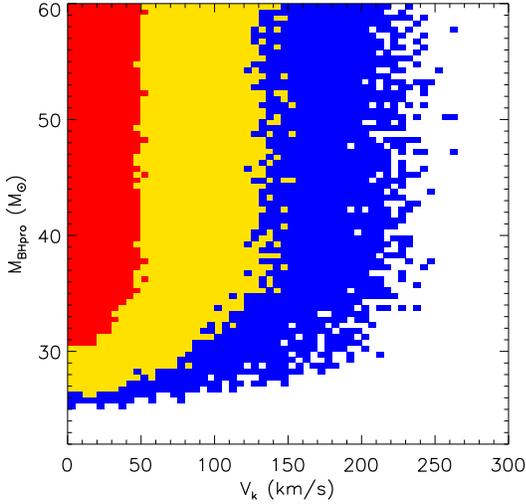}
\caption{Two dimensional joint $V_k$--$M_{BHpro}$ confidence levels: $68.3\%$ (red), $95.4\%$ (yellow), and $99.7\%$ (blue).}
\label{fig:Vk-Mbhpro-2dcl}
\end{center}
\end{figure}

We notice that the $M_{BHpro}$ PDF contains a plateau between 38 and 60~$M_\sun$, which indicates there is no clear upper limit on $M_{BHpro}$ beside the expected maximum
value based on the stellar evolutionary theory. Fortunately, our derived upper limit on $V_k$ depends very weakly on our adopted value of maximum $M_{BHpro}$.
Figure~\ref{fig:Vk-Mbhpro-2dcl} illustrate the two dimensional joint $V_k$--$M_{BHpro}$ confidence levels. For $M_{BHpro}$ above 38~$M_\sun$, the boundaries of the confidence
levels are almost perpendicular to the $V_k$ axis. This means the $V_k$ PDF remains roughly the same even if we omit the data points in a certain mass range in this regime.
Hence, our derived upper limit on $V_k$ will not change significantly if we adopt a lower $M_{BHpro}$ upper limit.

\subsection{Common Envelope Evolution Constraints}
\label{sec:CE_constraints}

Soon after the formation of the BH, the companion star evolves off the main-sequence and expands rapidly. When it fills its Roche lobe at the periastron, its mass and radius are
$78^{+8}_{-7}~M_\sun$ and $1000^{+320}_{-380}~R_\sun$, respectively.
Meanwhile, the orbital separation at periastron ($A_{peri,preCE}$) is $2200^{+740}_{-800}~R_\sun$.
Then, the binary undergoes dynamically unstable mass transfer, which leads to CE evolution. The common envelope efficiency ($\alpha_{CE}$) is constrained to be $\ge 0.6$.
We note that our values of $\alpha_{CE}$ is similar to those determined from studies of white dwarf binaries \citep[see][]{nelemans-tout05, demarco...11, davis...12}.
However, a direct comparison is not appropriate, as both the properties of the examined systems and the assumptions in the analyses are significantly different.

At the end of CE event, the binary consists of a 25--39~$M_\sun$ BH and a 32--42~$M_\sun$ He star.
The orbit of this binary is assumed to be circular, with a radius constrained to be 17--22\;$R_\sun$.
The equivalent orbital period is 25--35.0\;hr.
Unlike other limits presented in this section, the limits on the post-CE binary properties enclose the full range of the derived constraints,
which are obtained from the post-CE binary modeling discussed in \S\ref{sec:postCE_binary_modeling}. 

\section{Conclusion}
\label{sec:conclusion}

In this analysis, we track the evolution of IC 10 X-1 backwards in time up to the instant just before the core-collapse event and study the formation of the BH in this system. This covers
the following evolutionary phases: binary orbital dynamics at core collapse, CE evolution, and evolution of the BH--He star binary progenitor of the observed system.
We first focus on the latter and use the modeling of binary evolution to construct successful evolutionary sequences, and use them to determine the post-CE binary
properties. These sequences are referred as successful because their properties at present matches these observational
constraints of IC 10 X-1: binary orbital period, component masses and luminosity of the WR star.
Our predicted X-ray luminosity at the present time, resulting from the stellar-wind accretion onto the BH, is consistent with the observed values.
We then analyze the evolution through the necessary CE phase. We employ different CE treatments, as the standard treatment leads to unphysical results. We find that only the ``enthalpy''
formalism \citep{ivanova-chaichenets11} along with an energy-based CE efficiency \citep{webbink84} can explain physically the post-CE binary properties of the IC 10 X-1 progenitor.
Finally, we perform a Monte Carlo simulation on the orbital dynamics involved in the core collapse event.
Each data point contains seven free parameters drawn from uniform and isotropic distributions, which describe the properties of the pre-SN binary and the natal kick imparted to the BH.
Besides the constraints related to the core collapse event, we also use what we learned about the CE event involved in the formation of IC 10 X-1 as additional constraints to reject data points.
If a data point satisfies all the constraints mentioned in \S\ref{sec:dynamics@CC}, such as the survival of the binary through the core collapse event and the common envelope
efficiency $\alpha_{CE}$ being $\le 1$, we classify it as a successful data point. Our constraints (at 95.4\% of confidence) related to the BH formation and the CE event occurred in the past of
IC 10 X-1 are derived from all successful data points.
We find that the BH immediate (He rich) progenitor has a mass of $46 \pm 14~M_\sun$ and constrain the magnitude of the natal kick imparted to the BH to be $\le 130$\,km/s.

From the formation studies of low mass BH X-ray binaries, envelope ejection of massive stars during CE evolution has long been know to be energetically problematic
\citep[e.g.][]{kalogera99, podsiadlowski...03, wiktorowicz...13}.
In this study, we adopt the energetic formalism of \cite{webbink84} to calculate the CE efficiency ($\alpha_{CE}$).
Using the original \cite{webbink84} prescription, we find that the binding energies ($E_{bind}$) of our He star progenitor models are at least 3 times larger than the
available orbital energy, leading to $\alpha_{CE} > 1$ (see \S\ref{sec:CE_trial1}).
As $E_{bind}$ of our He star progenitors can potentially be smaller by adopting a different core definition, we examine four other core definitions (see \S\ref{sec:CE_trial2}).
Surprisingly we find that, for the very massive supergiants relevant here, these definitions all give roughly the same $E_{bind}$.
This result is in contrast to what \cite{tauris-dewi01} found in their study of $E_{bind}$ for RGB and AGB, in which they argued that $E_{bind}$ can be varied significantly by
adopting a different core definition.
However, the RGB and AGB stars considered by \cite{tauris-dewi01} are relatively much-lower mass stars, with initial masses up to 20\;M$_\sun$.

We also look at alternative CE treatments.
We find that neither enhancing the convective core overshooting (\S\ref{sec:CE_trial2}) nor doubling the mass loss rates (\S\,\ref{sec:CE_massloss}) of our He star progenitor
models can sufficiently decrease $E_{bind}$ to make $\alpha_{CE} \le 1$.
We also consider the formalism of hyper-critical accretion onto the BH during CE evolution (see \S\,\ref{sec:CE_trial3}).
In order to have $\alpha_{CE} \le 1$, we find that the radius of the He star progenitor at the onset of the hyper-critical accretion phase needs to be at least 1.3 times larger than
that at the onset of CE phase, but there cannot be any pre-CE binary configuration satisfying this requirement.
The last CE treatment considered in our analysis is the ``enthalpy'' formalism \citep{ivanova-chaichenets11}.
We find that it naturally provides $\alpha_{CE} \le 1$ with realistic pre-CE binary configurations, because including the term $P(m)/\rho(m)$ as an energy source lowers $E_{bind}$
sufficiently.
By adopting this CE treatment and considering \emph{all} constraints on our derived evolutionary history of IC 10 X-1, we find $\alpha_{CE}$ to be in a range of 0.6--1
(at 95.4\% of confidence).

Whether this ``enthalpy'' formalism is physically justified is not yet a settled issue \citep[see][]{ivanova...13}.
There are other potential energy sources discussed in the literature, which were not explored in our analysis.  
\cite{ivanova02} and \cite{voss-tauris03} suggested that the energy released from accretion onto the compact object during CE evolution can contribute to envelope ejection.
\cite{soker04} argued that this type of accretion will produce jets that help disperse the envelope.
A detail discussion on all potential energy sources can be found in a recent review on CE evolution by \cite{ivanova...13}.

Alternatively, there are uncertainties in building our He star progenitor models that can lead to a decrease in $E_{bind}$.
As mentioned in \S\ref{sec:methodology}, our models did not account for the effects of rotation.
Rotational mixing \citep{maeder-meynet00, maeder09book} can increase the mass of the He core and change the internal structure of the H rich envelope.
Also, we did not include rotationally enhanced mass loss rates, which can decrease the mass of the envelope.
Furthermore, although we find that doubling the mass loss rates does not significantly lower the $E_{bind}$ of our models (see \S\,\ref{sec:CE_massloss}),
the uncertainties in the mass loss rates of massive stars beyond the main-sequence phase are larger than a factor of 2.
For instance, \cite{petrovic...05} allowed a factor of 6 uncertainty in their mass loss rates for WR stars.
Given the uncertainties in modeling the evolution of massive stars, we cannot rule out the possibility that by fine-tuning our He star progenitor models we can decrease $E_{bind}$ sufficiently
and obtain $\alpha_{CE} \le 1$ without invoking the contribution of extra energy sources during CE evolution.

In this study, we uncover the evolution of IC\;10\;X-1 back to the point just prior to the formation of the BH.
Although we did not extend our detailed binary modeling further backwards in time, we can illustrate one scenario of how a primordial binary evolves to the
current state of IC\;10\;X-1, which is based on the results of one successful data point in our analysis.
We start with a binary consisting of two zero age main sequence stars, which are $\sim$ 150 and $\sim 87\;M_\sun$.
The binary orbit is initially circular, with an orbital radius of $\sim 8300\;R_\sun$ (equivalent to a period of $\sim 16$\;yr).
This orbit is so wide that throughout the evolution of the primary, its radius is at least four times smaller than its Roche lobe.
Hence the orbital evolution is dominated by mass loss from the system via stellar wind.
Soon after the more massive primary evolves off the main sequence, it loses its H rich envelope due to its massive stellar wind and becomes a He rich star.
This He star also suffers from intensive mass loss.
Just before collapsing to a BH, the mass of this He star is $\sim 44\;M_\sun$.
At the same time, the secondary approaches the end of its main sequence evolution, with a core H abundance of $\sim 9\%$.
Its mass and radius are $\sim 82\;M_\sun$ and $\sim 40\;R_\sun$, respectively.
Due to huge mass loss from the system, the orbital period increases to $\sim 55$\;yr.
During the core-collapse event, the BH (He rich) immediate progenitor loses $\sim 11\;M_\sun$ and forms a $\sim 33\;M_\sun$ BH.
Meanwhile, a natal kick of $\sim 70$\;km/s is imparted to that BH.
Right after the core-collapse event, the binary orbit is very eccentric ($e \approx 0.85$) and the orbital period increases to $\sim 64$\;yr.
In $\lesssim 0.3$\;Myr, the secondary becomes a supergiant and fills its Roche lobe at periastron, leading to CE evolution.
At the end of CE evolution, the binary consists of a $\sim 33\;M_\sun$ BH and a $\sim 35\;M_\sun$ He star in a circular orbit, with an orbital period of $\sim 34$\;hr.
This binary continues to evolve to the current state of IC\;10\;X-1.
We emphasize that we do not necessarily consider this scenario as unique;
instead we use it as the simplest possible scenario that can be consistent with the evolution of a primordial binary.

Based on our derived evolutionary history of IC 10 X-1, the spin angular momentum of the BH immediate progenitor is likely to be low.
This is because once it loses its H rich envelope, the BH progenitor will suffer from the high mass loss rates of WR stars.
This intense mass loss via stellar wind will take most angular momentum away from the BH progenitor and spin it down quickly.
Tidal effects could have kept the the BH progenitor from spinning down. However, the pre-SN binary orbit is relatively wide, with an orbital period of $\ge 0.5$~yr.
This means that the tides exerted on the BH immediate progenitor by its companion star are expected to be weak.
Even if the tidal interactions are much stronger than our expectation, they can at best synchronize the spin of the BH immediate progenitor with the orbital frequency.
Hence, under the assumption that spin angular momentum is conserved during the core collapse event, we argue that the natal spin of the BH in IC 10 X-1 is likely to be small.
Furthermore, since the BH has accreted negligible amount of mass from the stellar wind of its companion, it cannot be significantly spun up after its formation.
Therefore, we expect the current spin of the BH in IC 10 X-1 to be small as well. 

The BH spin of IC 10 X-1 is expected to be measured in the coming years. If the spin of this BH turns out to be fairly high, it may imply that the BH was spun up by accreting
significant amount of mass during CE evolution at a hyper-critical accretion rate.
Another possible explanation could be that the BH was spun up during the core collapse event \citep[see][]{blondin-mezzacappa07}.

\acknowledgements
This work has been primarily supported by the NSF Grant AST-0908930 awarded to V.K.;
T.W. also acknowledges partial support through the CIC/Smithsonian pre-doctoral fellowship program and the hospitality of the CfA;
T.F. acknowledges support from the CfA and the ITC prize fellowship programs.
E.G. acknowledges support from the NWO under grant 639.041.129.
The calculations for this work were performed on the Northwestern University Fugu cluster, which was partially funded by NSF MRI grant PHY-0619274,
and on the Northwestern University Quest High Performance Computing cluster.

%%%%%%%%%%%%%%%%%%
%%% Beginning of Appendix  %%%
%%%%%%%%%%%%%%%%%%
\appendix

\section{Hyper-Critical Accretion onto Black Hole During Common Envelope Evolution}

Let us denote the mass of the black hole (BH) by $M_{BH}$, the mass of the BH companion and its core mass by $M_{com}$ and $M_{core}$, and the binary semi-major axis
by $A$. Accretion onto the BH will be initiated when the binary semi-major axis equals to the BH companion's radius. As the time interval between the onset of common envelope (CE)
evolution and that of the accretion phase is relatively short, the masses of the BH and its companion are expected to be unchanged. Because of the very short circularization
time scale due to the Roche lobe filling BH companion, the binary orbit is assumed to be circular at the onset of accretion. CE evolution and accretion onto the BH will end when the envelope
of the BH companion is ejected to infinity.

Following \cite{belczynski...02}, the rate of change in $M_{BH}$ and $A$ with respect to $M_{com}$ due to accretion onto the BH is given by:
\begin{equation}
\left [ c_d (M_{com} + M_{BH}) - M_{com} \right ] \frac{dM_{BH}}{dM_{com}} = -\frac{M_{BH} M_{com}}{A}\frac{dA}{dM_{com}} + M_{BH},
\label{eqn:diffeqn1}
\end{equation}
where $c_d$ is the drag coefficient of the BH traveling in its companion's envelope. We adopt $c_d$ to be 6 \citep{shima...85, bethe-brown98}. In order to express these rates in
ordinary parameters, we use the equation of energy balance during CE evolution.

During the phase of accretion, the binding energy of the BH companion's envelope can be expressed as
\begin{align}
E_{bind} & = f_{ej} \cdot \frac{G M_{com} (M_{com} - M_{core})}{\lambda A} \nonumber \\
                 & = \frac{G M_{com} (M_{com} - M_{core} - \Delta M_{BH})}{\lambda A},
\label{eqn:Ebind2}
\end{align}
where $f_{ej}$ is the mass fraction of the ejected envelope as defined in Equation~(\ref{eqn:ejectedMF}) and $\Delta M_{BH}$ is the amount of mass accreted onto the BH.
Since the outcome of the CE event is known, which is given by the successful evolutionary sequences, we express $\Delta M_{BH}$ using the known post-CE BH mass ($M_{BH,f}$)
\begin{equation}
\Delta M_{BH} = M_{BH,f} - M_{BH},
\label{eqn:dMbh}
\end{equation}
By substituting Equation~(\ref{eqn:dMbh}) into Equation~(\ref{eqn:Ebind2}), we obtain
\begin{equation}
E_{bind} =  \frac{G M_{com} (M_{com} - M_{core} - M_{BH,f} + M_{BH})}{\lambda A}.
\label{eqn:Ebind3}
\end{equation}
The parameter $\lambda$ is a numerical factor scaling the binding energy of the BH companion's envelope during the phase of accretion, and is defined as
\begin{equation}
\lambda \equiv \frac{G M_{com,i} (M_{com,i} - M_{core})}{E_{bind,i} R_{com,i}}.
\end{equation}
Here, $M_{com,i}$ and $R_{com,i}$ are the pre-CE mass and radius the BH companion, respectively. We notice that $R_{com,i}$ is different from the BH companion's radius at
the onset of the accretion phase ($R_{com,acc}$). The envelope binding energy $E_{bind,i}$ of the pre-CE BH companion is defined in Equation~(\ref{eqn:Ebind}).

The envelope of the BH companion is ejected at the expense of the binary orbital energy. According to Equation~(\ref{eqn:mod_ceerg}), we can write
\begin{equation}
\alpha_{CE} \cdot \frac{dE_{orb}}{dM_{com}} = \frac{dE_{bind}}{dM_{com}}
\label{eqn:ceErgBal}
\end{equation}
Here, the rate of change in orbital energy with respect to $M_{com}$ is given by
\begin{equation}
\frac{dE_{orb}}{dM_{com}} =
-\frac{G}{2A} \left (
M_{com}\frac{dM_{BH}}{dM_{com}}
- \frac{M_{BH}M_{com}}{A}\frac{dA}{dM_{com}}
+ M_{BH}
\right )
\end{equation}
From Equation~(\ref{eqn:Ebind3}), we derive the rate of change in $E_{bind}$ with respect to $M_{com}$ as
\begin{equation}
\frac{dE_{bind}}{dM_{com}} =
\frac{G}{\lambda A}
\left [ M_{com}\frac{dM_{BH}}{dM_{com}}
+ \frac{ M_{com}(M_{core}-M_{com} + M_{bh,f} - M_{BH})}{A}\frac{dA}{dM_{com}}
+ 2M_{com}-M_{core} - M_{bh,f} + M_{bh}
\right ]
\label{eqn:dEbind3}
\end{equation}
Using Equation~(\ref{eqn:ceErgBal})--(\ref{eqn:dEbind3}), we can write
\begin{align}
\left ( 1 + \frac{2}{\alpha_{CE} \lambda} \right ) & M_{com} \frac{dM_{BH}}{dM_{com}} \nonumber \\
= \frac{M_{com}}{A} & \left [ M_{BH} - \frac{2}{\alpha_{CE} \lambda} (M_{core}-M_{com}+M_{BH,f}-M_{BH}) \right ] \frac{dA}{dM_{com}}
- M_{BH}
- \frac{2}{\alpha_{CE} \lambda} ( 2M_{com} - M_{core} - M_{BH,f} + M_{BH} )
\label{eqn:diffeqn2}
\end{align}
Then, we can derive a system of two ordinary differential equations using Equation (\ref{eqn:diffeqn1}) and (\ref{eqn:diffeqn2}):
\begin{align}
\frac{dA}{dM_{com}} & = \frac{A}{M_{com}} \left ( 1 + \frac{2 M_{com}}{\alpha_{CE} \lambda h_2} \right ) \label{eqn:dA_dM}\\
\frac{dM_{BH}}{dM_{com}} & = -\frac{2 M_{BH} M_{com}}{\alpha_{CE} \lambda h_1 h_2} \label{eqn:dMbh_dM}
\end{align}
where
\begin{align}
h_1 & = c_d (M_{BH} + M_{com}) - M_{com}\\
h_2 & = \frac{M_{BH} M_{com}}{h_1} \left ( 1 + \frac{2}{\alpha_{CE} \lambda} \right ) + M_{BH} + \frac{2}{\alpha_{CE} \lambda} (M_{com} - M_{core} - M_{BH,f} + M_{BH})
\end{align}

Using the known post-CE parameters as initial conditions, we can obtain the BH mass ($M_{BH,i}$) and semi-major axis ($A_{acc}$) at the onset of the accretion phase
by numerically integrating Equation~(\ref{eqn:dA_dM}) and (\ref{eqn:dMbh_dM}) from $M_{core}$ to $M_{com,i}$. In order to calculate the pre-CE semi-major axis ($A_{i}$),
we again use the energy balance equation $\alpha_{CE} \Delta E_{orb} = \Delta E_{bind}$. Here
\begin{equation}
\Delta E_{orb} = \frac{1}{2}G M_{BH,i} M_{com,i} \left ( \frac{1}{A_{acc}} - \frac{1}{A_i} \right )
\end{equation}
and
\begin{equation}
\Delta E_{bind} = \frac{G M_{com,i} (M_{com,i}-M_{core})}{\lambda} \left ( \frac{1}{R_{com,i}} - \frac{1}{R_{com,acc}} \right )
\end{equation}
Since the radius of the BH companion equals to the binary semi-major axis at the onset of the accretion phase, we set $R_{com,acc}~=~A_{acc}$ and obtain
\begin{equation}
A_i = \left [ 1 + \frac{2 (M_{com,i} - M_{core})}{\alpha_{CE} \lambda M_{BH,i}} \left ( 1 - \frac{A_{acc}}{R_{com,i}}\right ) \right ]^{-1} A_{acc}
\label{eqn:hyp_aprece}
\end{equation}
As the binary orbit continues to shrink during CE evolution, $A_i > A_{acc}$. Also, the BH companion must completely fill its Roche lobe in the pre-CE binary orbit.
Hence, $A_i$ needs to satisfy the constraint
\begin{equation}
R_{com,i} = A_i (1-e_i)  r_L
\end{equation} 
with a pre-CE orbital eccentricity ($e_i$) $< 1$. Here, $r_L$ is the approximated Roche lobe radius \citep{eggleton83}.

\end{document}